\documentclass[twocolumn,prl]{revtex4}
\usepackage{graphics,graphicx}
\begin{document}
\title{Spin Dynamics of the Spin-1/2 Kagom\'e Lattice Antiferromagnet ZnCu$_{3}$(OH)$_{6}$Cl$_{2}$}
\author{J.S.~Helton$^{1}$}
\author{K.~Matan$^{1}$}
\author{M.P.~Shores$^{2}$}
\author{E.A.~Nytko$^{2}$}
\author{B.M.~Bartlett$^{2}$}
\author{Y.~Yoshida$^{3}$}
\author{Y.~Takano$^{3}$}
\author{A.~Suslov$^{4}$}
\author{Y.~Qiu$^{5}$}
\author{J.-H.~Chung$^{5}$}
\author{D.G.~Nocera$^{2}$}
\author{Y.S.~Lee$^{1*}$}
 \affiliation{$^{1}$Department of Physics, Massachusetts Institute of Technology, Cambridge, MA 02139}
 \affiliation{$^{2}$Department of Chemistry, Massachusetts Institute of Technology, Cambridge, MA 02139}
 \affiliation{$^{3}$Department of Physics, University of Florida, Gainesville, FL 32611}
 \affiliation{$^{4}$National High Magnetic Field Laboratory, Tallahassee, FL 32310}
 \affiliation{$^{5}$NIST Center for Neutron Research, Gaithersburg, MD 20899 and Department of Materials Science and Engineering, University of
Maryland, College Park, MD, 20742 \\
$^*$email: younglee@mit.edu}
\date{\today}

\begin{abstract}

We have performed thermodynamic and neutron scattering measurements
on the $S=1/2$ kagom\'e lattice antiferromagnet
ZnCu$_{3}$(OH)$_{6}$Cl$_{2}$. The susceptibility indicates a
Curie-Weiss temperature of $\theta_{CW} \simeq -300$~K; however, no
magnetic order is observed down to 50~mK. Inelastic neutron
scattering reveals a spectrum of low energy spin excitations with no
observable gap down to 0.1~meV. The specific heat at low-$T$ follows
a power law temperature dependence.  These results suggest that an
unusual spin liquid state with essentially gapless excitations is
realized in this kagom\'e lattice system.

\end{abstract}

%\pacs{75.40.Gb, 78.70.Nx, 75.40.-s}
\maketitle

An important challenge in condensed matter physics is the search for
quantum disordered ground states in two dimensional systems. Of
particular interest is studying quantum spin liquids, an example of
which is the ``resonating valence bond'' state proposed by
Anderson\cite{Anderson2}.  These states are unusual in that neither
translational nor spin rotational symmetries are broken. It is
believed that the $S=1/2$ Heisenberg antiferromagnet on a kagom\'e
lattice (composed of corner sharing triangles) is an ideal system to
look for spin liquid physics due to the high degree of frustration.
There is broad theoretical consensus that the ground state of the
$S=1/2$ kagom\'e antiferromagnet is not magnetically
ordered\cite{MisguichLhuillier,Elser,Marston,Singh,Sachdev,Waldtmann,Wang}.
However, many basic properties are still under debate, such as the
magnitude of the gap to the first triplet state.  An intriguing
possibility is the existence of deconfined $S=1/2$ spinons as the
fundamental excitations, as opposed to conventional $S=1$ magnons.

Despite heavy theoretical interest, experimental studies of the
$S=1/2$ kagom\'e lattice have been hampered by the difficulty in
synthesizing such materials.  Here, we report thermodynamic and
neutron scattering measurements on powder samples of
ZnCu$_{3}$(OH)$_{6}$Cl$_{2}$, known as
herbertsmithite\cite{Braithwaite}. As has been previously
reported\cite{Shores}, Zn$_{x}$Cu$_{4-x}$(OH)$_{6}$Cl$_{2}$ can be
synthesized with variable Zn concentration, from $x$=0 to $x$=1
(herbertsmithite).   Figure 1(a) represents the transformation from
Cu$_{2}$(OH)$_{3}$Cl, which has a distorted pyrochlore structure, to
ZnCu$_{3}$(OH)$_{6}$Cl$_{2}$, which consists of Cu kagom\'e layers
separated by nonmagnetic Zn layers.  Structurally,
ZnCu$_{3}$(OH)$_{6}$Cl$_{2}$, with space group $R\bar{3}m$ and
lattice parameters $a=b=6.832$~\AA~and $c=14.049$~\AA, appears to be
an excellent realization of the $S=1/2$ kagom\'e lattice
antiferromagnet.  Initial evidence is the absence of long-range
magnetic order, as shown in the neutron diffraction scans in
Fig.~1(b). In Cu$_{2}$(OH)$_{3}$Cl, clear magnetic Bragg peaks are
observed below $\sim6$~K; whereas no magnetic Bragg peaks are
observable down to 1.8 K in ZnCu$_{3}$(OH)$_{6}$Cl$_{2}$.

\begin{figure}
\centering
\includegraphics[width=7.2cm]{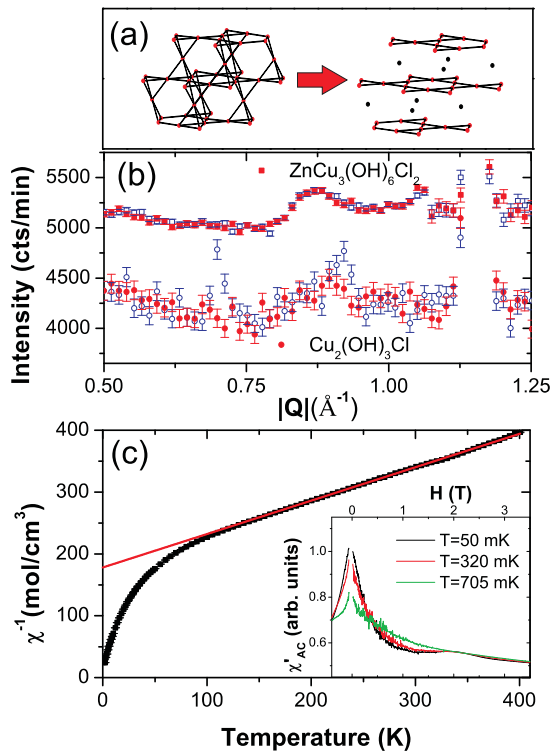} \vspace{-4mm}
\caption{(color online) (a) The chemical transformation from the
pyrochlorelike lattice of Cu$_{2}$(OH)$_{3}$Cl to the kagom\'e
layers of ZnCu$_{3}$(OH)$_{6}$Cl$_{2}$. (b) Magnetic diffraction
scans of the two systems at $T=1.4$~K (open) and 20~K (filled).  The
Cu$_{2}$(OH)$_{3}$Cl data show magnetic Bragg peaks at $Q\simeq0.70$
and $Q\simeq0.92$ which are absent for the
ZnCu$_{3}$(OH)$_{6}$Cl$_{2}$ data (which have been shifted by 2300
cts/min for clarity).  (c) Magnetic susceptibility of
ZnCu$_{3}$(OH)$_{6}$Cl$_{2}$ measured using a SQUID magnetometer
plotted as $1/\chi$, where mole refers to a formula unit. The line
denotes a Curie-Weiss fit. Inset: ac susceptibility (at 654 Hz) at
low temperatures measured at the NHMFL in Tallahassee, FL. }
\vspace{-4mm}
\end{figure}

To further characterize the properties of
ZnCu$_{3}$(OH)$_{6}$Cl$_{2}$, we performed magnetic susceptibility
measurements on powder samples.  The susceptibility, shown in
Fig.~1(c), can be fit to a Curie-Weiss law at high temperatures
($T>200$~K). The resulting Curie-Weiss temperature of $-300 \pm
20$~K implies an antiferromagnetic exchange $J\simeq17$~meV,
calculated using the series expansion corrections for the kagom\'e
lattice\cite{Harris,Grohol,Matan}.  The susceptibility continually
increases as the temperature is lowered down to 1.8~K. At first
glance, this behavior may suggest the presence of several percent
free spin-1/2 impurities yielding a Curie tail. This is certainly
possible, but is not necessarily the case.  From the chemical
analyses, we calculate the stoichiometric coefficients to be $3.00
\pm 0.04$ on the Cu site and $1.00 \pm 0.04$ on the Zn site. Also,
we have measured the ac susceptibility at temperatures down to
50~mK, as shown in the inset of Fig.~1(c). These data do not follow
the simple Brillouin function behavior expected for free $S=1/2$
spins. In particular, the susceptibility increase from 705~mK to
50~mK is much smaller than the free spin prediction. Recently, Ofer
and coworkers\cite{Keren} have shown that the muon Knight shift and
transverse relaxation rate have $T$ dependences similar to the
measured susceptibility.  Hence, the measured susceptibility may be
intrinsic to the Cu kagom\'e system.  We note that similar behavior
is found for the frustrated $S=1/2$ nuclear moments of $^3$He films
on graphite, where the susceptibility is found to continually
increase with decreasing temperature down to $T\sim
J/300$\cite{Masutomi}. Another recent $\mu$SR study\cite{Mendels}
emphasizes the role of defects.  The roles of impurities and
exchange or Dzyaloshinskii-Moriya\cite{Rigol} anisotropies in this
system remain important topics for further investigation. We also
observe a small peak in the ac susceptibility near $H=2$~T at 50~mK
which disappears upon warming to 705~mK.  The overall susceptibility
data indicate the absence of magnetic order or a spin gap down to
50~mK.

\begin{figure}
\centering
\includegraphics[width=7.2cm]{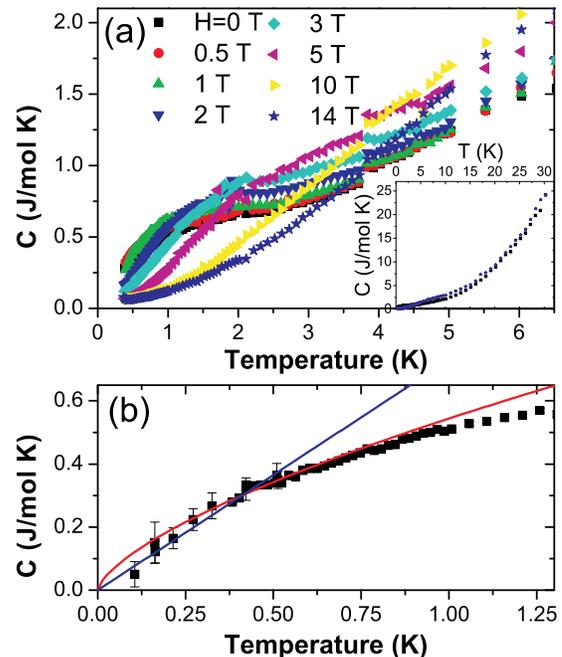} \vspace{-4mm}
\caption{(color online) (a) The specific heat $C(T)$ of
ZnCu$_{3}$(OH)$_{6}$Cl$_{2}$ in various applied fields, measured
using a Physical Properties Measurement System. Inset: $C(T)$
plotted over a wider temperature range in applied fields of 0 T
(square) and 14 T (star). (b) $C(T)$ in zero field measured down to
106~mK. The lines represent power law fits as described in the text.
} \vspace{-4mm}
\end{figure}

The specific heat $C(T)$ of ZnCu$_{3}$(OH)$_{6}$Cl$_{2}$ is shown in
Fig.~2(a) in various applied fields. For temperatures of a few
Kelvin and higher, the lattice contribution to the specific heat
(proportional to $\sim T^3$) is the most significant contribution,
as shown in the inset. However this contribution diminishes at low
temperatures, and below $\sim 5$~K, an additional contribution is
clearly observed which arises from the Cu spin system.  Magnetic
fields of a few Tesla can significantly affect the low-$T$ behavior,
and fields of 10 Tesla and higher strongly suppress the specific
heat below 3~K.  The difficulty in synthesizing an isostructural
nonmagnetic compound makes it hard to subtract the lattice
contribution precisely. However, the magnetic field dependence
suggests that the specific heat in zero applied field below 1 K is
predominately magnetic in origin.  As a rough measure of the spin
entropy, the field-induced change in specific heat below 3~K,
obtained by subtracting the 14 T data from the zero field data,
accounts for about 5\% of the total entropy of the spin system.

Additional specific heat measurements at zero field at temperatures
down to 106 mK were performed at the National High Magnetic Field
Laboratory (NHMFL) and the combined data are shown in Fig.~2(b). The
specific heat at low temperatures ($T<1$~K) appears to be governed
by a power law with an exponent which is less than or equal to 1. In
a 2D ordered magnet, magnon excitations would give $C \sim T^2$. The
kagom\'e-like compound SrCr$_{8-x}$Ga$_{4+x}$O$_{19}$
(SCGO)\cite{Ramirez} and other 2D frustrated magnets\cite{Nakatsuji}
are also observed to have $C \sim T^2$ even in the absence of
long-range order\cite{Sindzingre,MisguichBernu}. The behavior that
we observe in ZnCu$_{3}$(OH)$_{6}$Cl$_{2}$ below 1~K stands in
marked contrast. We can fit our data to the power law $C=\gamma
T^\alpha$, though we note that the exponent $\alpha$ is sensitive to
the chosen range of temperatures that are fit.  The blue line in
this figure represents a linear fit with $\alpha=1$ over the
temperature range $106~{\rm mK} < T < 400~{\rm mK}$.  The fitted
value for $\gamma$ is $240\pm20$ mJ/K$^2$ Cu mole.  If we include
higher temperatures, the red line represents a fit with $\alpha=2/3$
over the temperature range $106~{\rm mK} < T < 600~{\rm mK}$.
Extending the fitted range to even higher temperatures can yield
$\alpha$ values as low as 0.5.

Finally, inelastic neutron scattering measurements of the low energy
spin excitations were performed on deuterated powder samples of
ZnCu$_{3}$(OD)$_{6}$Cl$_{2}$.   High resolution measurements were
taken on the time-of-flight Disk Chopper Spectrometer (DCS) at the
NIST Center for Neutron Research in Gaithersburg, MD. A sample with
mass 9 g was cooled in a dilution refrigerator and studied with
incident neutrons of wavelength 7~{\AA}, yielding an instrumental
energy resolution of 0.02~meV (half-width at half-maximum).  As
shown in Fig.~3(a), the spin excitations form a broad spectrum at
low energies.  A notable observation is the near temperature
independence of the scattering for positive energy transfers.  The
excitation spectrum on the negative energy-transfer side is
suppressed at low temperatures due to detailed balance.

The magnetic scattering intensity is proportional to the dynamic
structure factor $S(\vec{Q},\omega)= (n(\omega)+1) \chi^{\prime
\prime} (\vec{Q},\omega)$, where $n(\omega)$ is the Bose occupation
factor and $\chi^{\prime \prime} (\vec{Q},\omega)$ is the imaginary
part of the dynamical susceptibility. We find that part of the
measured intensity for positive energy transfers below 0.4 meV is
spurious background scattering, probably caused by multiple
scattering of neutrons within the sample environment. Therefore, to
extract the intrinsic scattering from the sample, the following
procedure was used.  For negative energy transfers, $\chi^{\prime
\prime} (\omega,T{\rm = 10 K})$ can be obtained by subtracting the
35~mK data (which is essentially background) from the 10~K data and
dividing by the Bose factor. Here, $\chi^{\prime \prime} (\omega)$
represents the dynamical susceptibility integrated over momentum
transfers $0.25 \leq |\vec{Q}| \leq 1.5$~{\AA}$^{-1}$ and is a good
measure of the local response function.  This is plotted in
Fig.~3(b), where the positive $\omega$ data is obtained by using the
fact that $\chi^{\prime \prime} (\omega)$ is an odd function of
$\omega$. Then, $\chi^{\prime \prime} (\omega,T{\rm = 35 mK})$ can
be extracted from the positive energy transfer data using
$S(\omega;T{\rm = 35 mK})-S(\omega;T{\rm = 10 K})=I(\omega;T{\rm =
35 mK})-I(\omega;T{\rm = 10 K})$, where $I(\omega)$ is the measured
intensity and the background is assumed to be temperature
independent between 35~mK and 10~K.  As seen in Fig.~3(b), the data
for $\chi^{\prime \prime} (\omega)$ at $T=35$~mK increase with
decreasing $\omega$, indicating the absence of a spin gap down to
0.1 meV.  Moreover, the data may be described by a simple power law;
the solid line represents a fit to the form $\chi^{\prime \prime}
(\omega) \propto \omega^{\gamma}$ with an exponent
$\gamma=-0.7\pm0.3$. This apparently divergent behavior is unusual
and again differs markedly from measurement on SCGO\cite{Broholm}
which yield $\gamma \simeq 0$. Of course, within the errors, we
cannot rule out other functional forms for $\chi^{\prime \prime}
(\omega)$.

\begin{figure}
\centering
\includegraphics[width=7.2cm]{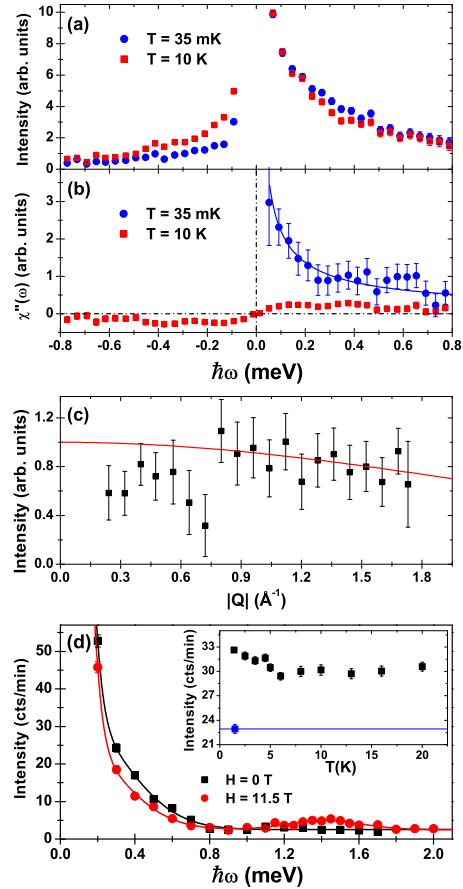} \vspace{-6mm}
\caption{(color online) (a) Inelastic neutron scattering data taken
on DCS, integrated over momentum transfers $0.25 \leq |\vec{Q}| \leq
1.5$~{\AA}$^{-1}$. (b) $\chi^{\prime \prime} (\omega)$, extracted
from the data as described in the text.  The line denotes a power
law fit. (c) The $Q$-dependence of the scattering, integrated over
energy transfers $-0.5 \leq \hbar\omega \leq -0.22$~meV. (d) Energy
scans taken on SPINS at zero field and 11.5 T at
$|\vec{Q}|=0.6$~\AA$^{-1}$ and $T=1.2$~K. The lines are guides to
the eye. Inset: Temperature dependence of the scattering for $0.3
\leq \hbar\omega \leq 0.5$~meV and $|\vec{Q}|=0.9$~\AA$^{-1}$.  The
blue data point and line indicate the background, measured on the
energy loss side at $T=1.5$~K.
 } \vspace{-4mm}
\end{figure}

The $Q$-dependence of the scattering is shown in Fig.~3(c). These
data were obtained by integrating over energy transfers $-0.5 \leq
\hbar\omega \leq -0.22$~meV and subtracting the 35~mK data set from
the 10~K data set.  We find that the data appear to be only weakly
dependent on $|\vec{Q}|$. Note that due to the polycrystalline form
of the sample, the data represents the powder average of the
scattering from a crystal. The solid line represents the squared
form factor $|F|^2$ for the Cu$^{2+}$ ion. The deviations of the
data from $|F|^2$ suggest that the structure factor is not
completely independent of $|\vec{Q}|$. That is, some degree of spin
correlations are necessary to account for the relative reduction in
scattering at small $|\vec{Q}|$. The overall diffuse nature of the
scattering points to the absence of a well-defined length scale to
describe these correlations.

Further measurements were taken using the triple-axis SPINS
spectrometer at the NCNR with the sample mounted inside a
superconducting magnet, as shown in Fig.~3(d).  The instrument was
configured in a horizontally focusing analyzer geometry with $E_f=
3.05$~meV and collimations of guide-$80^{\prime}$-radial-open. A BeO
filter was placed in the scattered beam to reduce higher-order
neutron contamination. The resulting instrumental energy resolution
was about 0.06~meV.  An applied field of 11.5 Tesla transfers
spectral weight from lower to higher energies. This demonstrates
that a significant fraction of the low energy scattering is magnetic
in origin, since the incoherent and phonon background would not
respond to the applied field in this manner. The magnetic signal in
zero field extends down to below 0.2 meV, consistent with the
analysis of the above DCS data. In 11.5 T, the magnetic signal
becomes peaked around $\hbar\omega \simeq 1.4$~meV, which is close
to the Zeeman energy $g\mu_B H$. However, the half-width of this
peak of about 0.21~meV is significantly broader than the resolution.
Therefore, the peak does not simply originate from Zeeman
excitations of noninteracting spins, which would result in a narrow
energy peak, but involve spins which are part of the interacting
system.  The integrated spectral weight of the zero field magnetic
signal for $\hbar\omega < 1$~meV accounts for at most 20\% of the
total scattering expected from a $S=1/2$ spin system (an estimate
made by normalization to the incoherent elastic scattering from the
sample and also to a vanadium standard). The inset of Fig.~3(d)
shows the temperature dependence of the inelastic signal with energy
transfers integrated over the range $0.3 \leq \hbar\omega \leq
0.5$~meV. There is a small increase in the signal when the sample is
cooled below $\sim 5$~K, though, for the most part, the intensity is
largely independent of temperature in this range.

Our experimental results suggest an intriguing picture for the
ground state properties of the $S=1/2$ kagom\'e lattice
antiferromagnet.  A hallmark of the quantum spin liquid in 2D is the
existence of deconfined $S=1/2$ spinons as the fundamental magnetic
excitation.  A rich variety of spin liquid states have been
theoretically proposed in which the spinons can be described as
bosonic\cite{Sachdev,Wang,Senthil1}, fermionic\cite{Lee1,Motrunich},
or even as Dirac fermions\cite{Ran}. We note that several of these
theories are based on triangular lattice Hamiltonians, and they may
not have clear extensions to the kagom\'e lattice antiferromagnet.
Using a naive comparison to a generic model of fermionic spinons
with a Fermi surface, one would expect $C=\gamma T$. From our linear
fit below 400 mK, the value of $\gamma$ indicates a Fermi
temperature of $T_F \sim 110$~K. However, other forms for the
specific heat (such as $C \propto T^2$) may hold at higher
temperatures where the lattice contribution prevents us from clearly
identifying the magnetic contribution.

The neutron scattering measurements of the excitation spectrum at
low temperatures are also consistent with expectations of deconfined
spinons in a spin liquid. We find no evidence of a spin gap down to
$\sim J/170$, much lower than the prediction from exact
diagonalization studies for a spin gap of $\sim
J/20$\cite{Waldtmann}. The power law behavior of $\chi^{\prime
\prime} (\omega)$ is interesting and may indicate a spin liquid with
critical spin correlations\cite{Altshuler}. Our observation of a
diffuse $Q$ dependence for the inelastic scattering suggests that if
a singlet spin liquid picture is correct, then the singlets are not
restricted to nearest neighbor dimers, since no well-defined length
scale is indicated by the data. The near temperature independence of
$S(\vec{Q},\omega)$, similar to observations in $f$-electron
systems\cite{Aronson}, may indicate the proximity to a quantum
critical point.  Many of the current theories for 2D spin liquids
were formulated to describe experimental results\cite{Coldea,Kanoda}
for $S=1/2$ triangular lattice systems.  More theoretical studies
based explicitly on the $S=1/2$ kagom\'e Heisenberg antiferromagnet
(including the possible effects of impurities and exchange or
Dzyaloshinskii-Moriya anisotropies) are certainly important for
further comparisons with experimental results.

We thank P.A. Lee, A. Keren, J.W. Lynn, Q. Huang, T. Senthil, and
X.-G. Wen for useful discussions and E. Palm and T. Murphy for help
with the measurements at the NHMFL. The work at MIT was supported by
the NSF under Grant No.~DMR 0239377, and in part by the MRSEC
program under Grant No.~DMR 02-13282. This work used facilities
supported in part by the NSF under Agreement No.~DMR-0454672.  A
portion of this work was performed at the NHMFL, which is supported
by NSF Cooperative Agreement No.~DMR-0084173, by the State of
Florida, and by the DOE.

\bibliography{kagome_spin_liquid}
\end{document}